\documentclass[sigconf,nonacm]{acmart}

\usepackage{multirow}

\newcounter{finding}
\newsavebox{\rqfindbox}

\newenvironment{rqblock}{%
  \par\medskip\noindent
  \begin{lrbox}{\rqfindbox}%
  \begin{minipage}{\dimexpr\linewidth-2\fboxsep-2\fboxrule\relax}%
  \textbf{Finding~\stepcounter{finding}\thefinding.}%
}{%
  \end{minipage}%
  \end{lrbox}%
  \fcolorbox{black!30}{white}{\usebox{\rqfindbox}}%
  \par\medskip
}

\begin{document}

%\title{Understanding Crate Hallucination in LLM-Generated Rust Code}
\title{When LLMs Invent Rust Crates: An Empirical Study of Hallucination Patterns and Mitigation}

% Authors (acmart style)
\author{Jieming Zheng}
\affiliation{%
  \institution{Southern University of Science and Technology}
  \city{Shenzhen}
  \country{China}
}
\email{12432728@mail.sustech.edu.cn}

\author{Hao Guan}
\affiliation{%
  \institution{Nankai University}
  \city{Tianjin}
  \country{China}
}
\email{hao.guan@nankai.edu.cn}

\author{Yepang Liu}
\authornote{Corresponding author. Jieming Zheng and Yepang Liu are affiliated with Research Institute of Trustworthy Autonomous Systems and Department of Computer Science and Engineering at SUSTech.}
\affiliation{%
  \institution{Southern University of Science and Technology}
  \city{Shenzhen}
  \country{China}
}
\email{liuyp1@sustech.edu.cn}

\begin{abstract}
% Insert your abstract here from abstract.tex
Large Language Models (LLMs) have become powerful tools for code generation, yet they remain prone to hallucinations—producing plausible but incorrect or fabricated outputs.
Among these, package hallucination, where an LLM suggests non-existent dependencies, poses an emerging security risk to the software supply chain.
While previous studies focus on popular languages like Python or JavaScript, in this work we present the first large-scale empirical study on crate hallucination in LLM-generated Rust code.
We construct a multi-source dataset combining coding tasks from \textsc{Stack Overflow}, \textsc{GitHub}, and LLM-generated tasks, and evaluate both commercial and open-source models under various decoding settings.
Our analysis reveals that, unlike prior findings in Python and JavaScript, hallucination behavior in Rust follows a distinct pattern: different models exhibit surprisingly consistent hallucination rates, and these rates show minimal sensitivity to model parameters.
Furthermore, we investigate prompt engineering strategies to mitigate hallucinations without sacrificing code quality.
This study provides new insights into the reliability and security implications of LLM-assisted Rust development, offering guidance for future research and safer model deployment in software engineering workflows.

\end{abstract}




% \date{Received: date / Accepted: date}

\maketitle

% \subclass{68N01 \and 68N15 \and 68N30}
% Software engineering, Programming languages, Software tools

\label{intro}
\section{Introduction}
With the rapid rise of generative AI, Large Language Models (LLMs) have been widely adopted for code generation~\cite{sengar2025generative,gozalo2023chatgpt,jiang2024survey}.
Programmers now rely on LLMs more than ever for coding.
Automated code generation powered by LLMs can boost developer productivity and reduce avoidable human errors during software development~\cite{peng2023impact,pandey2024transforming}.
However, \textbf{hallucination} remains a critical drawback of LLMs, resulting in outputs that appear plausible but are factually incorrect, logically unsound, or entirely fabricated~\cite{huang2025survey}.

In the specific context of code generation, Liu et al.~\cite{liu2024exploring} recently grouped hallucinations into three main categories: 1) intent conflicting, 2) context deviation, and 3) knowledge conflicting.
Their study further shows that code-generation hallucinations are widespread across various popular LLMs (e.g., \textsc{ChatGPT}, \textsc{CodeRL}, \textsc{CodeGen}).
Such hallucinations can have important reliability and security implications: fabricated or misleading code, including incorrect APIs and dependency-related suggestions, may introduce build failures, vulnerable implementation choices, or downstream opportunities for adversarial exploitation.
As a result, hallucination-driven attacks have emerged as a new and urgent topic in software supply chain security research~\cite{spracklen2025package,krishna2025importing,trendmicro2025slopsquatting}.

A critical weakness of code-generating LLMs is \textit{package hallucination}: the model recommends or imports a dependency that does not actually exist~\cite{spracklen2025package}.
Spracklen et al.~\cite{spracklen2025package} proposed a threat model based on package hallucination, in which an adversary can pre-register a malicious package under a hallucinated identifier so that users who follow the LLM-generated recommendation may inadvertently introduce a supply-chain compromise.

In this paper, we also focus on studying package hallucination within the context of code generation.
Existing research on package hallucination primarily focuses on the most popular languages (e.g., Python, JavaScript), while paying little attention to less popular languages like Rust~\cite{spracklen2025package,krishna2025importing}.
Rust is typically adopted for system-level programming, including operating systems, browsers, and blockchain protocols, where security is of paramount importance. 

The threat model introduced by prior work~\cite{spracklen2025package} is compelling, but it manifests differently in Rust.
In Rust, downloading an external crate requires an explicit dependency declaration in \texttt{Cargo.toml}.  Referencing that crate in \texttt{.rs} source files alone does not trigger its download.
Therefore, a hallucinated crate reference in Rust code should not be interpreted as direct evidence that a malicious package would be fetched or installed.
Even so, it is still practically important to study such hallucinated crate references.
They are a direct reliability problem, since invalid crate references can lead to unresolved imports, incorrect namespace usage, and build failures.
In addition, many LLM-assisted coding workflows generate source-code snippets, functions, or single-file examples rather than complete Rust projects with fully specified \texttt{Cargo.toml} manifests.
As a result, source-level crate references often serve as the earliest observable signal and an upstream precursor of dependency-related confusion, while dependency-resolution decisions occur only later through searches on \emph{crates.io}, package-management tools, or manual crate additions.
Motivated by this distinction, our study focuses on characterizing crate hallucinations at the source-code reference level in Rust.

The lack of research on LLM hallucinations in Rust code generation may lead to the following consequences: (1) we cannot determine whether the hallucination patterns observed in other language ecosystems also manifest in Rust; (2) we lack evidence to discover new hallucination modes that may arise from Rust’s unique scenarios; and (3) we are unable to design or evaluate mitigation tools that are truly effective for Rust developers.

To bridge this research gap, we present the first large-scale evaluation of crate hallucinations of Rust code generated by both commercial and open-source LLMs.
In response to the lack of prior work on Rust, we propose the following research questions (RQs):

\begin{comment}
\begin{itemize}
	\item \textbf{RQ1a (Cross-model):} Do different LLM architectures differ significantly in crate-level hallucination rate?
	\item \textbf{RQ1b (Within-family):} Within a single model family, how does parameter count (size) affect hallucination?
	\item \textbf{RQ2:} Within the same LLM, do decoding settings (temperature, top-\(p\), max tokens, etc.) affect the hallucination rate of Rust crate recommendations?
	\item \textbf{RQ3:} What types of hallucinations occur in LLM-generated Rust crate recommendations, and what common patterns do they share?
	\item \textbf{RQ4:} Does the hallucination rate vary across Rust application domains?
	\item \textbf{RQ5:} Which post-processing or prompt-engineering strategies can significantly reduce hallucination rates in Rust crate recommendations?
\end{itemize}
\end{comment}

\begin{itemize}

	\item \textbf{RQ1 (Prevalence of Crate Hallucinations)}: How likely are LLMs to hallucinate when recommending Rust crates? How do model family, model size, and decoding parameters affect this likelihood?

	\item \textbf{RQ2 (Common Hallucination Patterns)}: What recurring patterns can be observed in Rust crate hallucinations, and what shared characteristics do these errors exhibit?

	\item \textbf{RQ3 (Mitigation of Crate Hallucinations)}: Can we leverage common prompt engineering strategies to mitigate Rust crate hallucination?

\end{itemize}

These questions aim to fill the gap in understanding crate hallucinations in Rust, offering insights into model behavior, parameter sensitivity, error classification, domain-specific variations, and mitigation strategies.
To answer them, we studied crate hallucinations on 14 models from six model families using a multi-source Rust coding task dataset containing 2,794 coding tasks.
	Our study yielded interesting findings.
	Regarding prevalence (RQ1), we found that crate hallucinations remain consistent across model families, model sizes, and decoding parameters.
	% and are only weakly affected by decoding settings (RQ1).
	From these crate hallucinations, we identified four recurring, Rust-specific hallucination patterns that suggest structured confusions (RQ2).
	Finally, we found that common mitigation strategies at the prompting stage, such as RAG and self-refinement, cannot effectively reduce the likelihood of crate hallucination (RQ3), suggesting the need for more sophisticated techniques to prevent crate hallucination.

In summary, we make the following contributions in this paper.
\begin{itemize}
	\item To the best of our knowledge, we are the first to conduct a systematic empirical study on crate hallucination in LLM-generated Rust code.
	\item We conducted our empirical study on a large-scale coding task dataset, with a wide range of large language models.
	\item We identified four recurring patterns of crate hallucination and found common prompting-stage mitigations ineffective.
	\item We constructed a dataset containing 2,794 Rust coding tasks.
	      We release this dataset to the public to facilitate future research in this domain.
	       % The dataset is available at https://figshare.com/s/f19fe17afb291ad386fa.
	      % \item \textbf{We provide a systematic empirical characterization of crate hallucinations in LLM-generated Rust code.} We evaluate the prevalence of crate hallucinations across a diverse set of tasks and models, and examine how the risk varies with model family, model size, and decoding configurations to establish an evidence-based understanding of when and how often the issue occurs.
	      % \item \textbf{We build a reusable dataset for large-scale study of Rust crate hallucinations.} We create a coding task dataset from multiple sources, enabling consistent identification and measurement of hallucinated crates.
	      % \item \textbf{We conduct an in-depth analysis of Rust-specific crate hallucination patterns and highlight key error modes.} We uncover recurring Rust-specific hallucinations and show that many cases arise from consistent confusion with real crates rather than random invention.
	      % \item \textbf{We evaluate lightweight mitigation strategies and derive practical guidance for reducing crate hallucinations.} We test prompt-level improvements without modifying model weights, compare their effectiveness and stability, and summarize what works in practice for controlling supply chain risk when generating Rust code.
\end{itemize}

\section{Background}
\label{sec:background}

\par\textbf{Hallucinations in Code Generation by LLMs.} LLM hallucination is the phenomenon where a model generates unfaithful or nonsensical content~\cite{ji2023survey}.
Hallucinations are commonly categorized into three types: (1) \textit{input-conflicting} hallucinations, which deviate from the user-provided input; (2) \textit{context-conflicting} hallucinations, which contradict the model’s own earlier outputs; and (3) \textit{fact-conflicting} hallucinations, which are inconsistent with established world knowledge or cannot be verified~\cite{zhang2025siren}.
Moreover, this phenomenon is typically driven by data, training, and inference~\cite{huang2025survey}.

In code generation scenarios, hallucinations manifest as syntactically well-formed or semantically plausible code that relies on fabricated, non-existent, or contextually misaligned elements, making them difficult to detect through manual inspection alone~\cite{gao2025systematic,tian2025codehalu}.
In programming, these hallucinations are often categorized as: (1) \textit{intent conflicting}, where the generated code deviates from the user’s task requirements and shows low semantic relevance; (2) \textit{context deviation}, where the code misaligns with the given context or the model's own earlier outputs; and (3) \textit{knowledge conflicting}, where the output violates domain facts or programming conventions~\cite{liu2024exploring}.
Recently, the growing popularity of ``vibe coding'' has encouraged more users to rely on LLMs to generate and iteratively refine code from natural-language intents~\cite{ray2025review}.
However, multiple empirical studies have shown that even advanced coding assistants still hallucinate~\cite{liu2023your, zhang2025siren, zhang2025llm}.
As a result, this problem has attracted increasing attention, motivating a rapidly expanding body of work on systematically characterizing code hallucinations and developing detection and mitigation techniques to improve the reliability of LLM-generated code~\cite{gao2025systematic}.

\par\textbf{Package Hallucination.} Package hallucination is a type of knowledge-conflicting hallucination in code generation.
It refers to scenarios where an LLM recommends non-existent or incorrect package names while generating code~\cite{spracklen2025package}.
This issue is particularly noteworthy because dependency suggestions often appear plausible, leading users who trust LLM-generated code to directly install the package if it becomes available.
Recent studies have shown that such hallucinated package names are highly prevalent in LLM-generated code, and this phenomenon spans different model types and programming languages~\cite{spracklen2025package,krishna2025importing}.
Beyond correctness, package hallucination introduces a concrete supply chain threat: attackers can register a package under a hallucinated name with malicious code, turning an LLM’s fabricated recommendation into an installation-time attack.
Such an attack is called a \textit{package confusion attack}~\cite{neupane2023beyond}.
Given that many users may utilize LLM-generated code without careful review, this creates substantial opportunities for such attacks~\cite{al2025ai, ryser2025calibrated}.
Given its pervasiveness, cross-model impact, and tangible supply chain risks amplified by developers’ unchecked reliance on LLM outputs, package hallucination emerges as a critical and understudied problem that demands systematic research to mitigate its threats to software security and reliability.

\par \textbf{Rust.}
Rust is a systems programming language designed to provide strong memory-safety guarantees without relying on a garbage collector, largely enabled by its ownership and borrowing model~\cite{rustbook_stable}.
These features have made Rust as a high-performance substrate for modern developer tooling, leading to the emergence of various Rust-based alternatives within established ecosystems.
For example, projects like Polars~\cite{polars_docs} and uv~\cite{uv_github} for Python, as well as Deno~\cite{deno_github} and SWC~\cite{swc_github} for JavaScript, have gained widespread popularity by delivering significantly higher performance than legacy implementations.
This growing reliance on Rust-based infrastructure underscores the importance of the security and integrity of its library ecosystem.

In practice, Rust development heavily relies on Cargo and the \emph{crates.io} ecosystem, where functionality reuse is commonly achieved by adding third-party crates.
At the language level, a crate is a unit of compilation and distribution and contains a tree of nested module scopes.
This crate–module structure and Rust’s path-based naming conventions interact closely with dependency-related hallucinations.
First, Rust distinguishes crates (external dependency units) from modules (in-crate namespaces), yet both may appear in similar path-based forms in code, which can foster structured confusions such as treating standard library modules as external crates~\cite{rustbook_stable}.
Second, adding a dependency is not purely declarative: Cargo compiles and executes \texttt{build.rs} as part of the build process, and procedural macros run at compile time over Rust syntax, so adding an unintended crate can have security implications even before the program runs~\cite{rustblog2025_malicious_crates_fasterlog_asyncprintln}.
More broadly, recent studies show that LLMs can hallucinate packages or crates during code generation and that adversaries may exploit hallucinated or confusingly similar names to mount supply chain attacks.
The Rust ecosystem has also seen official reports of malicious crates~\cite{rustblog2025_malicious_crates_fasterlog_asyncprintln}, underscoring the practical relevance of this security risk.
%\\yepang{This paragraph is a bit long and contains much content. See if we can split it into two paragraphs.} \jm{revised}

% Crate ecosystem, hallucination, threat model.

\label{sec:approach}
\section{Data Collection}
In this section, we detail our approach to collecting and constructing the datasets utilized in our study.
Additionally, we discuss the rationale for selecting the subject models.

\subsection{Collecting Coding Tasks}
To effectively evaluate package hallucinations in Rust code generation, we need a coding task dataset that offers broad coverage of programming tasks, Rust-specific contexts, and scenarios where the use of external crates is natural or recommended.
To this end, we constructed a large-scale task dataset from three different data sources.
Our dataset encompasses a wide range of programming scenarios, capturing both real-world developer needs and synthetically generated tasks.

\par\textbf{\textsc{Stack Overflow} Coding Tasks.}
\textsc{Stack Overflow}~\cite{stackoverflow_website} is a prominent developer Q\&A platform where questions are tagged and ranked by community feedback.
We leverage this data source to obtain practical, user-driven Rust coding needs that reflect authentic developer inquiries and real-world challenges.
Rust-related questions are collected by querying posts tagged with \texttt{Rust} and ranking them according to community votes.
Our objective is to produce a global ranking of the most upvoted Rust questions.
This approach prioritizes widely validated, general-purpose questions that are likely to translate into practical, model-ready tasks.
To ensure focus and stability, we retain only unique question entries and remove duplicates across pages.
Applying this process yields 1,001 unique Rust questions, which constitute the \textsc{Stack Overflow} component of our coding task dataset.
We observe that not all questions retrieved from \textsc{Stack Overflow} by this method are strictly coding tasks.
Following Spracklen et al.'s work~\cite{spracklen2025package}, we do not apply further filtering to these questions, but instead instruct the LLM to produce code wherever possible.

\par\textbf{\textsc{GitHub} Coding Tasks.}
\textsc{GitHub}~\cite{github_website} hosts real-world open-source projects with rich code, comments, and usage context.
We leverage this platform to collect project-grounded tasks that reflect realistic developer needs and implementation details observed in production code.
To construct our dataset, we first query \textsc{GitHub} for Rust repositories and rank the results by the number of stars, selecting the top 100 projects.
However, we observe that many highly starred repositories are functionally similar (e.g., \textit{uv}~\cite{uv_github}, \textit{ruff}~\cite{ruff_github}, and \textit{ty}~\cite{ty_github}, all of which are Rust-implemented, high-performance Python developer tools).
To enrich the diversity of programming tasks, we apply random sampling and manual inspection to ensure the selected projects span a range of functionalities.
From this curated pool, we sample 10 repositories for downstream processing.

\begin{figure}[t]
	\centering
	\Description{A documented Rust function is transformed into a coding task by combining its RustDoc comments, function signature, and an implementation instruction.}
    \includegraphics[width=0.8\linewidth]{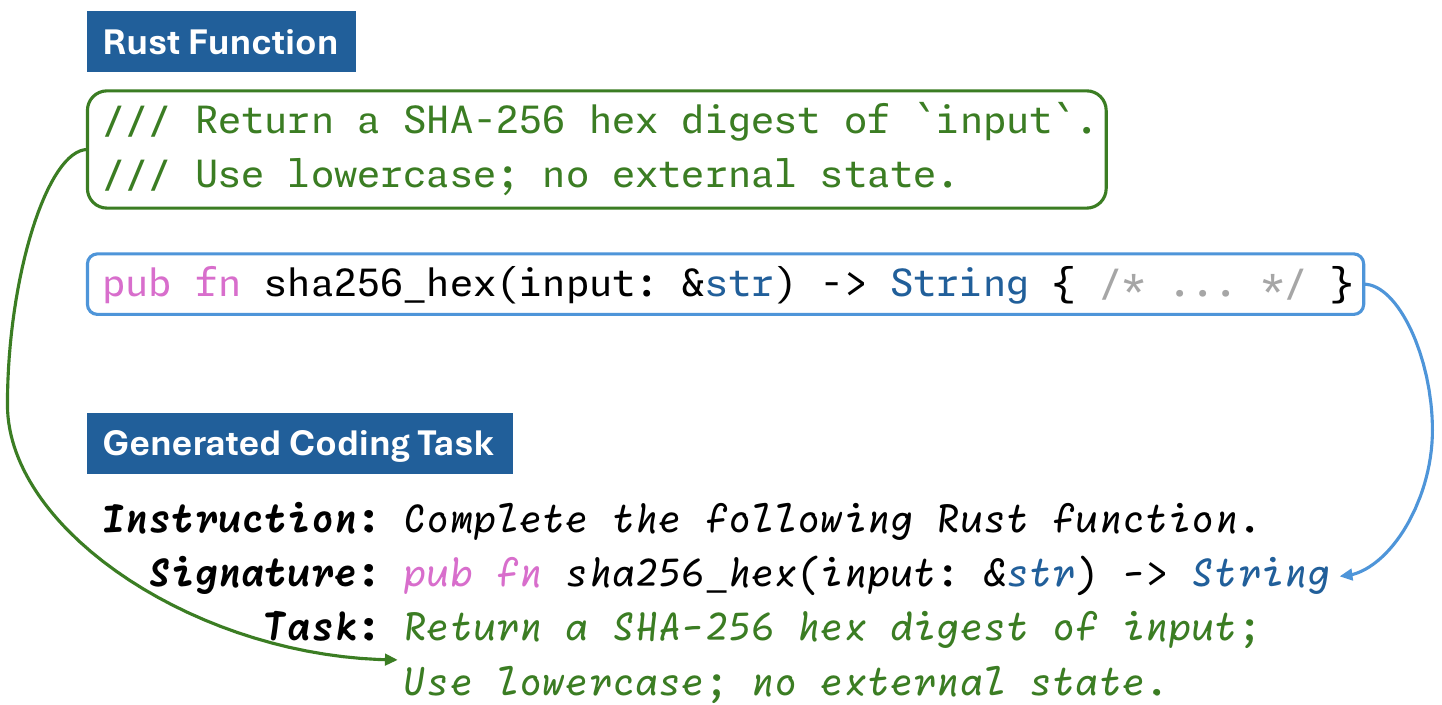}
	\caption{Converting a Rust Function into a Coding Task}
	\label{fig:task-generation}
\end{figure}

For each of these 10 repositories, we transform repository artifacts into coding tasks by extracting function-level documentation.
Specifically, we extracted all the function definitions from \texttt{.rs} files and selected those documented with RustDoc comments.
To create each coding task, we pair the extracted brief instruction with its corresponding function signature.
Figure~\ref{fig:task-generation} illustrates the conversion from a Rust function into a coding task.
For selected functions, we extracted function definitions with RustDoc comments from all Rust source files. Each function's RustDoc summary and signature form a coding task, ensuring tasks are grounded in real-world, well-specified code.
This yielded 1,011 coding tasks.
% \jmrev{To keep the task focused on the function itself, we intentionally omit file-level \texttt{use} imports and type aliases.}
% \jmrev{We retain comments even when they mention external crates, as prohibiting crate references would make the tasks less realistic and would not reflect how developers document Rust code in practice.}

To ensure the quality and generality of the coding tasks, we applied a second-pass, rule-based review to the comment-derived tasks with the following rules.
% \begin{enumerate}
	\par \textbf{Vague or underspecified task.}
	      These tasks lack sufficient detail, such as I/O behavior or clear semantics, making them unfit for general-purpose evaluation.
	      For example, RustDoc like ``Create a dummy node'' is too vague to be an actionable coding task.
	\par \textbf{Compiler internals.}
	      Tasks that are only meaningful within the compiler’s source-code context, often involving internal types, are excluded.
	\par \textbf{FFI/ABI or OS-specific constraints.}
	      Tasks that require platform-bound ABIs or system-specific behavior reduce portability and evaluability.
	      % For example, \lstinline{rust}{pub extern "msp430-interrupt" fn msp430()} where \texttt{msp430-interrupt} is a target-specific ABI, making it unsuitable for general code-generation evaluation.
	\par \textbf{Test directives (harness/build scripts).}
	      Tasks mixing compile flags or \textsc{FileCheck}-like directives do not represent implementable requirements.
	      For example, functions with ``\texttt{CHECK-NEXT:}'' comments are unsuitable for general code-generation evaluation.
% \end{enumerate}

Following these rules, we removed 218 items from the initial 1,011 coding tasks, leaving 793 for experiments.

\textbf{LLM-generated Coding Tasks.}
Many third-party libraries are created to solve a specific set of programming tasks, and their package descriptions explicitly summarize the provided functionality and typical use cases.
Therefore, package descriptions offer a practical source for constructing dependency-related coding tasks.
In particular, a crate description summarizes the capability the library provides and the scenarios it targets, which can be naturally rewritten into an implementation request.
This makes crate descriptions a scalable and consistent way to construct tasks where the expected solution requires introducing external dependencies.
Moreover, sampling crates from registries allows us to cover a broader range of different domains and dependency names.
Motivated by these considerations, we construct an LLM-generated coding task set by transforming the descriptions of selected crates into implementation-oriented tasks.

Inspired by the approach of Spracklen et al.~\cite{spracklen2025package}, which rewrites popular package descriptions into tasks, we construct a Rust LLM-generated coding task set as follows. To better reflect real-world usage, we select the popular crates from the popularity ranking on \emph{lib.rs}.
% \yepang{Why do we use underline?}.
However, since the popularity ranking on \emph{lib.rs} exposes only the top 250 crates, we adopt a mixed collection strategy: we take the top 250 crates from \emph{lib.rs} (Popular) and supplement them with 750 crates from \emph{crates.io}, which are ranked by recent downloads (deduplicating overlaps).
Subsequently, we employed the Qwen2.5:32B model to transform each crate description into exactly one coding task: an English sentence beginning with the fixed prefix ``Generate Rust code that \ldots''.
This procedure helped us produce 1,000 LLM-generated coding tasks.

\begin{table}[t]
	\centering
	\caption{Information of Evaluated Models}
	\label{tab:model_info}
	\small
	\begin{tabular}{@{}llcc@{}}
		\toprule
		\textbf{Model} & \textbf{Vendor} & \textbf{Model Size} & \textbf{\shortstack{Open-\\Source}} \\
		\midrule
		GPT-5 & OpenAI & Unknown & $\boldsymbol{\times}$ \\
		Gemini 2.5 Pro & Google & Unknown & $\boldsymbol{\times}$ \\
		Claude 4 Opus & Anthropic & Unknown & $\boldsymbol{\times}$ \\
		Qwen2.5-coder & Alibaba &
		\begin{tabular}[t]{@{}c@{}}0.5B, 1.5B, 3B,\\7B, 14B, 32B\end{tabular}
		& \checkmark \\
		DeepSeek-Coder-V2 & High-Flyer & 16B, 236B & \checkmark \\
		OpenCoder & INF Technology & 1.5B, 8B & \checkmark \\
		\bottomrule
	\end{tabular}
\end{table}

\subsection{Model Selection}
\label{sec:model_selection}
We evaluate crate hallucinations across multiple LLMs to understand whether the phenomenon is \textbf{model-dependent} and to improve the generalizability of our findings.
To cover both widely used proprietary assistants and reproducible open-source alternatives, we select models from two categories: commercial LLMs and open-source code LLMs.

We select models using public code-generation leaderboards as a practical criterion, aiming to evaluate crate hallucinations on strong coding models rather than on weak baselines.
Specifically, we rely on two widely used benchmarks for model selection: SWE-bench Verified~\cite{jimenez2023swe} for commercial assistants and EvalPlus~\cite{liu2023your} for open-source code LLMs.

SWE-bench Verified is a widely used benchmark for software-engineering tasks with verification-based scoring, and it is commonly used to compare mainstream proprietary assistants.
EvalPlus extends HumanEval-style evaluations with stronger test suites and stricter checking, and it has been widely adopted in recent work for comparing open-source code models~\cite{spracklen2025package, hui2024qwen2, lozhkov2024starcoder, huang2024opencoder}.

\par\textbf{Commercial Models.} Based on the SWE-bench Verified leaderboard, we choose Claude 4 Opus, GPT-5, and Gemini 2.5 Pro, which represent top-performing proprietary assistants from Anthropic, OpenAI, and Google, respectively.

\par\textbf{Open-source Models.} Based on the EvalPlus leaderboards, we choose Qwen2.5-Coder, DeepSeek-Coder-V2, and OpenCoder, which are among the top-performing open-source code models under this evaluation.

\label{sec:RQ1}
\section{RQ1: Prevalence of Crate Hallucinations}

In this section, we first describe how we detect crate hallucinations in LLM-generated Rust code, followed by the experimental setup and evaluation environment.
We then report the results for RQ1 and provide an analysis of how model family, model size, and decoding parameters relate to the likelihood of crate hallucinations.

\subsection{The Approach of RQ1}

RQ1 aims to measure how often LLMs hallucinate when recommending Rust crates and to compare hallucination rates across models.
Specifically, we evaluate the six models listed in Section~\ref{sec:model_selection} on three datasets.
All models are run with their default configurations.
For open-source models, we select the largest available parameter count.
To answer RQ1, for each model’s generated code, we apply the following hallucination-detection procedure.

\par\textbf{Detection of Crate Hallucinations.}
\label{detection_method}
To ensure accurate detection of crate hallucinations, we constructed a comprehensive inventory of existing Rust crates following the steps listed below.
\begin{itemize}
	\item We first scraped the complete list of crate names available on \emph{crates.io}, which is the official Rust package registry.
	\item Second, the Rust Standard library~\cite{rust_std_docs} consists of five built-in crates (\texttt{alloc, core, proc\_macro, std}, and \texttt{test}) that are not distributed via \emph{crates.io}.
	      We supplement the crate name list crawled in the first step with those five names obtained from the documentation.
	\item Next, we collected the set of \texttt{rustc} compiler crates (e.g. \texttt{mdman}, \texttt{rustc\_abi}) from the Rust compiler documentation~\cite{rustc_middle_docs}, as these crates can also be used in Rust programming.
	\item A further complication in the Rust ecosystem is that a package’s published name on \emph{crates.io} may differ from the crate name used in Rust programs.
	      Such aliases should be declared in the \texttt{Cargo.toml} file of a Rust project's repository.
	      To address this issue, we mined all repositories of our collected crates to extract those alternative names and merged them into our crate name list.
	      Because a small subset of packages either do not provide a valid \textsc{GitHub} repository link or the provided link does not work, our alias coverage may not be fully complete. Consequently, in our study, the true hallucination rate of the LLM-generated Rust code is likely slightly lower than the reported rate.
\end{itemize}

To extract crate names from LLM-generated Rust code, we rely on three reference patterns:
\begin{enumerate}
	\item \textbf{Use declarations} (\texttt{use crate\_name::*;}) \\
    \textit{Example:} \texttt{\textbf{use} serde::Serialize;}
%     \begin{lstlisting}[language=Rust,basicstyle=\ttfamily\small]
% use serde::Serialize;
% \end{lstlisting}
	\item \textbf{Direct path prefixes} (\texttt{crate\_name::module::item})\\
    \textit{Example:} \texttt{tokio::spawn;}
%     \begin{lstlisting}[language=Rust,basicstyle=\ttfamily\small]
% tokio::spawn;
% \end{lstlisting}
	\item \textbf{External crate declarations} (\texttt{extern crate crate\_name;}) \\
    \textit{Example:} \texttt{\textbf{extern crate} regex;}
%     \begin{lstlisting}[language=Rust,basicstyle=\ttfamily\small]
% extern crate regex;
% \end{lstlisting}
\end{enumerate}
\noindent We then compare the crates extracted by our parser with the list we constructed following the approach described above.

\par\textbf{Metrics.}
We introduce the following metric to quantify the degree of hallucination in a consistent and comparable way across prompts and models.
Our idea is straightforward.
Any extracted name that is absent from the crate name list built following the above process is labeled as a hallucinated crate.
Thus, the crate hallucination rate (CHR) for a model can be computed as follows.
$$
	\mathrm{CHR} = \frac{\text{Number of hallucinated crates}}{\text{Total number of extracted crates}}
$$

\par\textbf{Environment.}
All the experiments are carried out on a Linux server equipped with an AMD EPYC 9654 96-Core Processor, along with 8 NVIDIA RTX 6000 Ada GPUs and 768 GB of system memory.

\subsection{Empirical Study Results}

In the experiment, we generated a total of 16,764 code snippets and extracted 48,494 crate recommendations.
Of these recommended crates, 10,779 were identified as hallucinations, yielding an overall hallucination rate (CHR) of 22.23\%.

\begin{figure}[t]
	\centering
	\includegraphics[width=0.95\columnwidth]{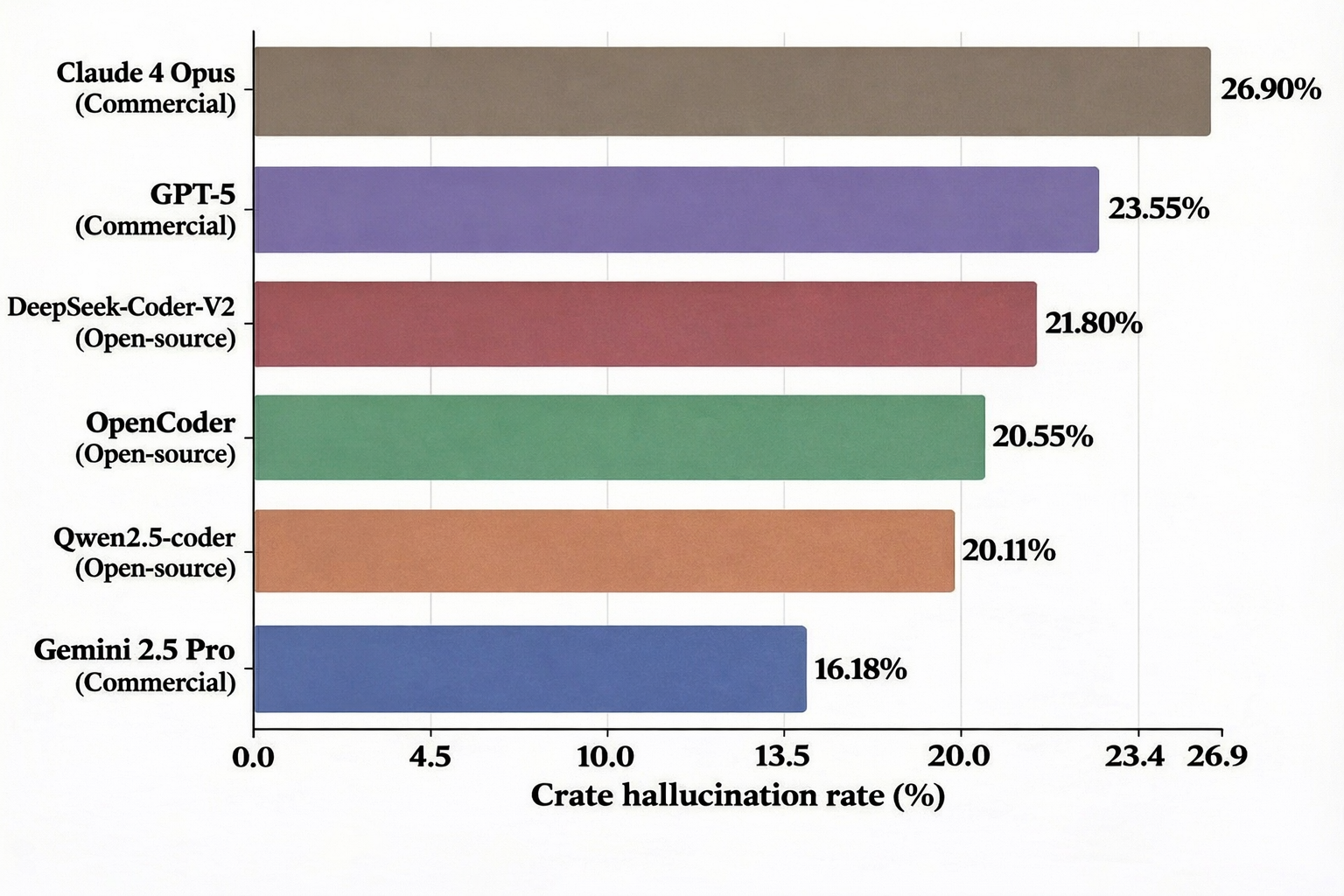}
	\caption{Crate Hallucination Rates of the Evaluated Models}
    \Description{A horizontal bar chart comparing crate hallucination
rates across six evaluated models. Gemini 2.5 Pro has the lowest
rate, while Claude 4 Opus has the highest rate.}
	\label{fig:rq1-a}
\end{figure}

%% Double-check the tenses here.
Overall, we observe that cross-model differences in CHR are not significant, though a few models achieve markedly lower CHR.
As shown in Fig. \ref{fig:rq1-a}, Gemini-2.5-Pro achieves an CHR of 16.18\% (the best) and Claude-4-opus's CHR is 26.90\% (the worst).
The ratio of worst to best is about 1.66, indicating that model choice is a very important factor.
Selecting suitable models can significantly boost the trustworthiness of the generated Rust code.

In the comparison between commercial and open-source models, the average overall CHR for commercial models is 23.34\% and that for open-source models is 20.79\%, indicating a slight advantage for open-source models.
Also, we found that Gemini-2.5-Pro is an outlier and excluding it yields a commercial macro-averaged CHR of 25.23\%, making the higher CHR of commercial models more apparent compared to open-source models.

\begin{rqblock}
	Crate hallucinations are widespread and observed across all the six selected LLMs.
	When generating Rust code, the overall differences in CHR between commercial and open-source models are not large; however, certain commercial models achieve notably lower CHRs.
\end{rqblock}

\textbf{The effect of model size.} In the second experiment, our goal is to quantify, within each LLM family, the relationship between model size (i.e., parameter count) and the crate hallucination rate, testing whether larger models would demonstrate lower CHRs.
Specifically, we evaluate the three open-source models listed in Section \ref{sec:model_selection}, along with all of their size variants available on Ollama (DeepSeek-Coder-V2: 16B/236B; OpenCoder: 1.5B/8B; Qwen2.5-Coder: 0.5B/1.5B/3B/7B/14B/32B).

As shown in Table \ref{tab:model_size_hallucination}, our experimental results indicate that model size does not correlate with reduced hallucination rates across the evaluated model families.
% \sout{For the Qwen2.5-Coder series, we computed the correlation coefficient between model size and hallucination rate, obtaining a value of $-0.059$, which suggests a negligible relationship.
% To formally assess this, we conducted a linear regression analysis to test whether the model size significantly predicts the hallucination rate.
% The null hypothesis ($H_0$) posits that the size of a model has no association with the model's hallucination rate, while the alternative hypothesis ($H_1$) posits an association.
% The p-value for the size coefficient is 0.912.
% Consequently, we fail to reject $H_0$ with $p < 0.05$, indicating that our data do not provide evidence for a statistically significant effect of model size on hallucination rate.}
Since Qwen2.5-Coder is the only evaluated family with more than two size variants, we further conduct an ANOVA-style F-test~\cite{fisher1970statistical} under a linear model on this family to assess whether model size explains the observed CHR variation.
The result shows no statistically significant effect of model size with $F(1,4)=0.0138, p=0.912$.
The corresponding effect size is also negligible, with $R^2=0.0034$.
These results indicate that, within the Qwen2.5-Coder family, parameter count explains almost none of the variation in CHR. 
\begin{table}[t]
	\centering
	\caption{Hallucination Rate by Model Size Across Different Model Families}
	\label{tab:model_size_hallucination}
    \small
	\begin{tabular*}{\linewidth}{@{\extracolsep{\fill}}lrr@{}}
		\toprule
		\textbf{Model~Family}              & \textbf{Model Size} & \textbf{Hallucination~Rate} \\
		\midrule
		\multirow{2}{*}{DeepSeek-Coder-V2} & 16B                 & 18.46\%                     \\
		                                   & 236B                & 21.92\%                     \\
		\midrule
		\multirow{2}{*}{OpenCoder}         & 1.5B                & 19.54\%                     \\
		                                   & 8B                  & 19.28\%                     \\
		\midrule
		\multirow{6}{*}{Qwen2.5-Coder}     & 0.5B                & 20.41\%                     \\
		                                   & 1.5B                & 21.62\%                     \\
		                                   & 3B                  & 18.78\%                     \\
		                                   & 7B                  & 20.83\%                     \\
		                                   & 14B                 & 21.36\%                     \\
		                                   & 32B                 & 20.07\%                     \\
		\bottomrule
	\end{tabular*}

\end{table}

We hypothesized that larger models would demonstrate lower CHRs; however, the result does not support this hypothesis.
Within the three families considered, model size is not a reliable predictor of hallucination rate in crate recommendation.
The effect of model size is small and statistically insignificant.
One plausible explanation is insufficient Rust coverage in pretraining or instruction data, which would limit the marginal benefit of additional parameters.
% However, our data do not directly measure corpus composition\yepang{what corpus?}, so we refrain from causal claims and instead outline tests to probe this explanation.

\begin{rqblock}
	For our tested open-source models, there is no statistically significant correlation between model size and crate hallucination rate.
\end{rqblock}

\begin{figure*}[t]
	\centering
	\includegraphics[width=\linewidth]{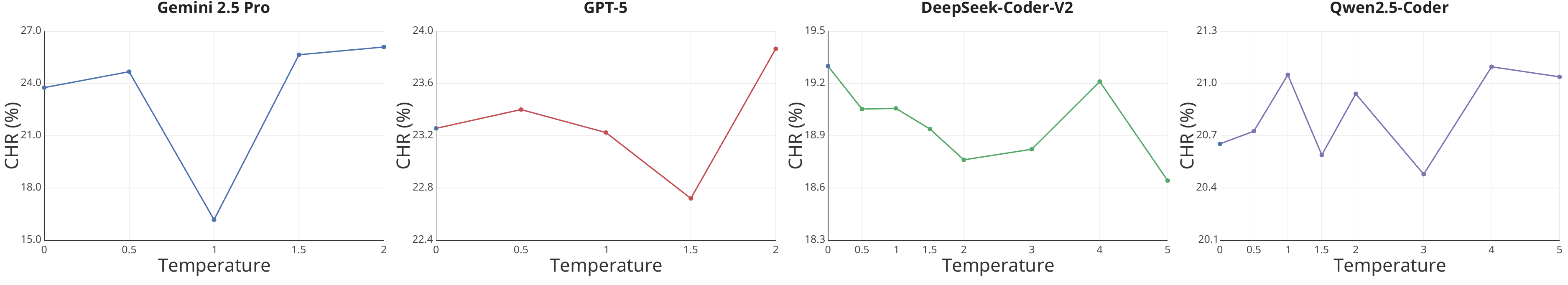}
	\caption{Hallucination Rates of Evaluated Models Across Different Temperature Settings}
    \Description{Four line charts showing crate hallucination rates at
different temperature settings for Gemini 2.5 Pro, GPT-5,
DeepSeek-Coder-V2, and Qwen2.5-Coder. The rates fluctuate without
a consistent increasing or decreasing trend.}
	\label{fig:temp_hr}
\end{figure*}

\textbf{The effect of
	temperature.}
We next investigate whether the sampling temperature parameter influences hallucination rate in Rust-crate recommendation.
Temperature controls how random the model’s output is: lower values give more deterministic, repeatable answers, while higher values encourage more diverse and creative generations.
Here, we hypothesize that increasing the decoding temperature will raise the CHR.
However, we observed that for Rust code generation, both open-source and commercial models show only marginal sensitivity to temperature.
As shown in Fig. \ref{fig:temp_hr}, the CHRs of all four models remain within a narrow range of 16.2\%–26.1\%, exhibiting small, inconsistent fluctuations rather than a clear upward trend.
For example, the CHR of Qwen2.5-Coder:32B fluctuates mildly between 20.65\% and 21.10\% as the temperature increases from 0 to 5, exhibiting no monotonic trend.
Among all models, Gemini-2.5-Pro shows a slightly wider fluctuation range, but the change is still modest and not statistically significant.

% \sout{To formally test this hypothesis, we performed a linear regression analysis to examine whether temperature significantly predicts hallu-\linebreak cination rate.
% The null hypothesis ($H_0$) posits that temperature has no association with hallucination rate, while the alternative hypothesis ($H_1$) posits an association.
% For Deepseek-Coder-V2, the p-value for the temperature coefficient is 0.109. 
% For Gpt5, the p-value is 0.713. 
% For Qwen2.5-coder, the p-value is 0.257. 
% For Gemini-2.5-pro, we observed a association with a p-value of 0.697.
% All four models showed no statistically significant association between temperature and hallucination rate.
% Namely, we fail to reject $H_0$ at conventional significance levels, indicating that our data do not provide evidence for a statistically significant effect of temperature on hallucination rate.}
To formally examine the effect of temperature, we conduct an ANOVA test using the temperature settings shared by all four evaluated models, i.e., 0, 0.5, 1, 1.5, and 2.
Since different models exhibit different baseline CHRs, we include model identity as a blocking factor and test the effect of temperature under the model CHR $\sim$ Model + Temperature.
The ANOVA result shows that temperature does not have a statistically significant effect on CHR with $F(4, 12) = 0.9159, p = 0.4859.$
In contrast, the model factor is statistically significant with 
$F(3, 12) = 5.0425, p = 0.0173$, which suggests that the observed variation is better explained by model differences than by temperature settings.
These results indicate that changing decoding temperature leads only to small and inconsistent fluctuations rather than a systematic effect on crate hallucination rate.
This observed stability contrasts with prior findings in Python and JavaScript~\cite{spracklen2025package}, where higher temperature often leads to increased hallucination rates.

\begin{rqblock}
	Unlike the findings in Python and JavaScript reported in previous work, for Rust, the correlation between temperature and crate hallucination rate is statistically weak.
\end{rqblock}

%rq1 summary

% In summary, our Rust experiments reveal a pattern that differs from prior work: hallucination rates are largely insensitive to model choice and decoding parameters.
% Across all models, the crate hallucination rate remains consistently around 20\%.

%\yepang{Please provide an explicit answer to RQ1 by summarizing the key findings. Please also do this for the other two RQs.}

\textbf{Answer to RQ1.} In summary, crate hallucination is prevalent in LLM-generated Rust code.
Across our experiments, the overall crate hallucination rate stays around 20\%, and all evaluated models exhibit non-trivial hallucinations.
Although the rate varies across models, we do not observe a consistent trend by model family or by whether the model is commercial or open-source. 
Within the open-source models we tested, increasing parameter count does not substantially reduce hallucination. 
Moreover, changing decoding temperature only leads to small and inconsistent differences. Overall, crate hallucinations in Rust appear relatively stable under model size and decoding-time adjustments.

\label{sec:RQ2}
\section{RQ2: Recurring Hallucination Patterns}

% RQ2 aims to examine whether crate hallucinations exhibit common patterns, and to analyze the shared characteristics underlying these recurrent errors.
% Because crate hallucinations can arise from different underlying causes, we analyze them from multiple angles to distinguish where they come from and how they recur.
% Analyzing module confusions helps isolate Rust-specific namespace mistakes from other hallucinations, while cross-model overlap indicates whether the errors reflect shared weaknesses rather than model-specific quirks.
% Examining name similarity further reveals whether hallucinations tend to be near-miss variants of real crates, which matters for practical risk mitigation design.
% Together, these analyses provide an interpretable basis for understanding common patterns and motivate our mitigation choices in RQ3.
%\yepang{This paragraph can be further revised to clearly enumerate the analyses we have done and their motivation.}
RQ2 aims to characterize recurring patterns in Rust crate hallucinations and to examine the shared characteristics of these errors. 
We conduct four complementary analyses that together reveal how these hallucinations are structured in practice:
% RQ2 aims to examine whether Rust crate hallucinations exhibit recurring patterns and what shared characteristics may underlie these errors.
% We conduct four targeted analyses to clarify the sources and recurring patterns of these hallucinations:
    \par \textbf{General analysis.} %We provide a high-level categorization of hallucinated references to obtain an overall picture of what types of errors occur and to identify common patterns that recur across coding tasks and models.
    We separate broad classes of hallucinated references to obtain an overall picture of what kinds of errors occur, especially whether the model is fabricating external crates or misusing existing Rust modules.
    \par \textbf{Cross-model analysis.} We measure how much hallucinated crate names overlap across different models, which helps distinguish systematic weaknesses from model-specific mistakes.
    \par \textbf{Name similarity analysis.} We examine lexical similarity between hallucinated names and real crates to test whether hallucinations tend to be near-miss variants, which is relevant for practical risk and mitigation design.
    \par \textbf{Cross-domain analysis.} We inspect frequent hallucinations for plausible transfer from other language ecosystems or platform APIs, helping explain how cross-ecosystem priors may shape erroneous Rust dependency suggestions.

\subsection{General Hallucination}
\label{subsec: GHA}
Our analysis reveals a critical finding: over half of the hallucinated crates are actually valid Rust modules that exist in the standard library or third-party crates, but are being misused or referenced in an unqualified way.
This suggests that the primary issue is not the generation of completely non-existent packages, but rather the improper usage of existing \textit{modules}.
A \textit{module} in Rust provides a hierarchical namespace that organizes and encapsulates code within a crate.
For example, ubiquitous data structures \texttt{HashMap} and \texttt{LinkedList} are encapsulated within the \texttt{std::collections} module.
The LLM-generated code may directly call \texttt{HashMap::new} without declaring \texttt{use std::collections::HashMap} in advance.
This issue of incorrect reference extends to
% the standard library (\texttt{std}) contains
numerous other standard library modules, including
%\texttt{std::collections},
\texttt{std::io}, \texttt{std::thread}, and \texttt{std::net}.
%\yepang{What exactly are the mistakes? How are the modules incorrectly referenced? Can we provide examples before presenting finding 4?}
%\gh{Added an example. Name the mistake as ``unqualified reference'' may be better.}
% \jm{revised by Hao Guan}
\begin{rqblock}
	Many hallucinated crate names are actually valid Rust modules that exist in the standard library, but are being incorrectly referenced.
\end{rqblock}

\begin{table}[t]
	\centering
	\caption{Top-10 Hallucinated Crate Names by All Models}
	\label{tab:top10_hallucinated_crates}
    \small
	\begin{tabular*}{\linewidth}{@{\extracolsep{\fill}}lr|lr@{}}
		\toprule
		\textbf{Crate}       & \textbf{Frequency} & \textbf{Crate}        & \textbf{Frequency}
		\\
		\midrule
		\texttt{thread}      & 1,086           & \texttt{trefcell}     & 296            \\
		\texttt{duration}    & 828            & \texttt{tu32}         & 282            \\
		\texttt{tokenstream} & 505            & \texttt{ttcplistener} & 277            \\
		\texttt{ordering}    & 418            & \texttt{thashset}     & 257            \\
		\texttt{vecdeque}    & 323            & \texttt{tpoll}        & 236            \\
		\bottomrule
	\end{tabular*}

\end{table}

Table~\ref{tab:top10_hallucinated_crates} %lists the ten most frequently hallucinated crates.
identifies the top ten hallucinated crates, ranked by their aggregate frequency of appearance within the generation for the coding tasks.
Notably, the majority of these names correspond to legitimate modules within the Rust standard library.
This indicates that nearly half of all hallucinations (45.47\%) are not entirely fabricated identifiers but rather cases where the model incorrectly treats an internal module as an external crate dependency.
Such errors suggest that the model partially recognizes semantically valid Rust entities, yet fails to represent their hierarchical position within the language’s namespace system.

This pattern suggests that LLMs may overgeneralize import conventions from languages such as Python, where modules and packages share similar import syntax, into Rust, which enforces a stricter separation between crates and modules.
This implication is consistent with our previous findings that many hallucinated crate names correspond to valid standard-library modules, indicating that the model captures their semantics but misrepresents their scope within the import hierarchy.

From a software supply chain perspective, although module-level hallucinations are less likely to result in malicious packages, they still pose challenges for automatic dependency management and build reliability.
Recent work also~\cite{ravi2025llmloop} showed that LLM-generated code often faces compilation errors and dependency issues.

\subsection{Hallucination Across Models}
Based on our prior analysis in Section~\ref{subsec: GHA}, we conducted a cross-model analysis separating hallucinations into module and non-module categories.
We compute pairwise Jaccard similarities between models for each category and quantify uniqueness by counting how many models produce each hallucination item.

We found that module hallucinations exhibit substantially higher cross-model overlap:  average Jaccard $\approx 0.44$ (max $\approx 0.50$).
Consistent with this, $55.05\%$ of module hallucinations are shared across models, while $44.95\%$ are unique.
In contrast, non-module hallucinations show much lower overlap—average Jaccard $\approx 0.16$ (max $\approx 0.21$)—and most cases are not shared: only $25.42\%$ are shared and $74.58\%$ are unique.

% These results suggest that errors involving module names are more systematic and generalizable—likely reflecting concentrated standard/common library namespaces and shared priors—thus yielding higher cross-model reproducibility
%\yepang{I cannot follow this sentence.}
%\gh{@jm Do you mean such errors may originate from the limitations/incompleteness of pre-training data?}
These results indicate that module-related errors are more consistent across models.
A likely reason is that such errors may originate from the incompleteness of pre-training data.
Thus, different models tend to make similar confusions. 
Conversely, non-module hallucinations are more context-dependent and model-specific.

\begin{rqblock}
	We found a moderate degree of overlap in crate hallucinations among models, with about half of module hallucinations being shared.
	This suggests that some hallucination patterns are common across different LLMs, though many remain model-specific.
\end{rqblock}

\begin{figure}[t]
	\centering
	\includegraphics[width=0.8\columnwidth]{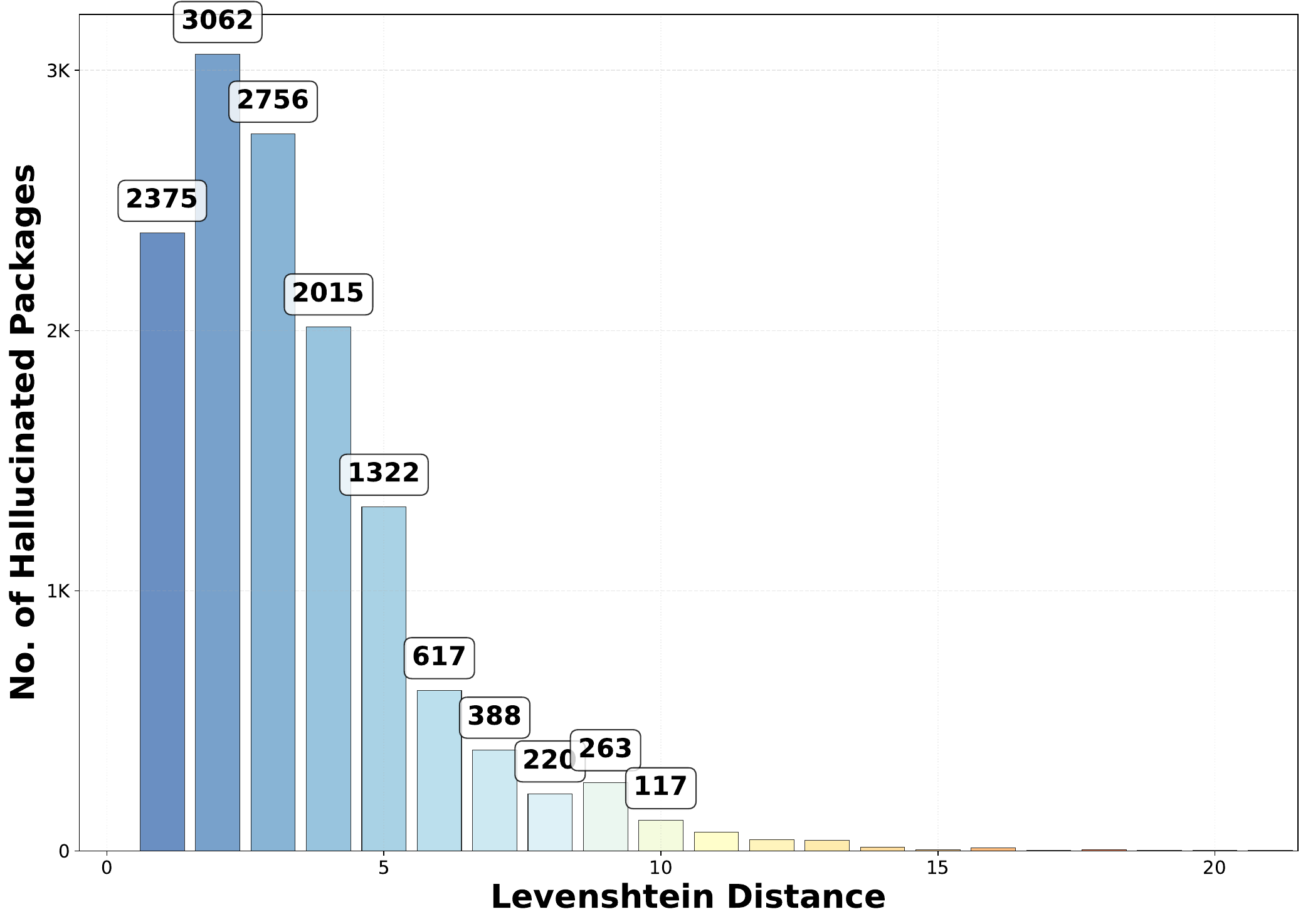}
	\caption{Distribution of Levenshtein Distance Between Hallucinated and Real Crate Names}
    \Description{A histogram showing the distribution of Levenshtein
distances between hallucinated crate names and their nearest real
crate names. Most occurrences are concentrated at small distances.}
	\label{fig:lev-dist}
\end{figure}

\subsection{Name Similarity of Hallucination}
To investigate the relationship between hallucinated crates and real crates in the Rust ecosystem, we employed the \textit{Levenshtein distance} \cite{lcvenshtcin1966binary} as a quantitative measure of string similarity.

As shown in Fig.~\ref{fig:lev-dist}, the results of the Levenshtein distance analysis indicate that hallucinated crate names in Rust are predominantly close variants of existing crates rather than entirely novel inventions.
Among all 5,647 unique hallucinated names remaining after excluding module-confusion cases (occurring 13,332 times in total), the weighted mean distance to the nearest valid crate is 3.42.
About 40.78\% of hallucination occurrences differ by only one or two characters, and another 45.70\% fall within a distance of three to five.
Only 2.36\% have a distance of ten or above, suggesting that most hallucinated crate names are minor variations involving limited character substitutions or omissions.
In total, 86.48\% of hallucination occurrences are within five edits of an existing crate, reflecting a high concentration of near-miss cases.

Further, we conducted a morphological analysis of these hallucinations to understand the underlying patterns and mechanisms of hallucination generation in the Rust ecosystem. The most prominent finding is the systematic inconsistency in hyphen usage.
Among 13,332 total cases, we observe that 32.96\% of hallucinated crates lack hyphens while their corresponding real packages contain hyphens.
This pattern suggests that the model tends to generate camelCase names, whereas the Rust ecosystem strongly favors kebab-case naming conventions.
Representative examples include:
\[
	\begin{array}{rcl}
		\texttt{httpresponse}   & \rightarrow & \texttt{http-response}   \\
		\texttt{tokentree}      & \rightarrow & \texttt{token-tree}      \\
		\texttt{servicebuilder} & \rightarrow & \texttt{service-builder} \\
		\texttt{genericarray}   & \rightarrow & \texttt{generic-array}
	\end{array}
\]

Another finding is that among close matches (edit distance $\leq$ 3, 8,193 occurrences), 4,879 cases (59.6\%) share common prefixes with their nearest legitimate counterparts, while 5,697 cases (69.5\%) exhibit suffix similarity.
This high degree of morphological overlap indicates that hallucinations are not randomly generated, but systematically constructed as variations of existing crate names.
Examples demonstrating this pattern include:
\[
	\begin{array}{lll}
		\texttt{downcast-ref} & \rightarrow & \texttt{downcast-rs} \quad \text{(prefix similarity),} \\
		\texttt{myenum}       & \rightarrow & \texttt{tyenum} \quad \text{(suffix similarity).}
	\end{array}
\]

Our findings contrast sharply with prior studies for Python and JavaScript~\cite{spracklen2025package}, which report that nearly half of hallucinations have edit distances of six or more, while our Rust analysis shows that 86.48\% of hallucinations fall within five edits of legitimate packages.
This discrepancy suggests that Rust hallucinations are more tightly clustered around existing names, potentially reflecting the more standardized naming conventions in the Rust ecosystem and the model's deeper internalization of Rust-specific patterns.

\begin{rqblock}
	Unlike prior findings for Python and JavaScript, for Rust, most hallucinated crate names are minor variations of real crates (with a few character edits).
	A prominent pattern is the misuse or omission of hyphens, with models often generating camel-case names instead of following the kebab-case convention used in Rust.
\end{rqblock}

\subsection{Hallucination Across Domains}
To identify common patterns among frequently occurring hallucinations, we analyze non-module hallucinations that occurred more than 10 times using a fine-grained cross-domain criterion: a hallucinated crate is considered ``cross-domain'' if its name clearly appears in other programming language ecosystems or system/platform ecosystems (e.g., Windows API, Android SDK).

Across the analyzed high-frequency hallucinated crates, we found that 38.9\% of them exhibit clear cross-domain characteristics.
For instance, ``\texttt{queryparser}'' is from the Python ecosystem and ``\texttt{winuser}'' is from Windows API.
This suggests that repeated hallucinations are not purely random, but often stem from exposure to overlapping semantics across ecosystems, indicating structured yet misaligned knowledge transfer.

\begin{rqblock}
	A substantial fraction of hallucinated crates reuse names from other ecosystems.
    % while remaining equally susceptible to the hyphenation errors described in Finding 6.
    % often with hyphenation errors similar to those seen in Rust-specific cases
\end{rqblock}

\textbf{Answer to RQ2.}
Crate hallucinations in Rust exhibit clear recurring patterns.
A large portion of errors are Rust-specific confusions where models generate unqualified references.
Hallucinated crate names also show systematic reuse and structure: many recur across different models, and a noticeable fraction appear as near-miss variants of real crates, indicating that models often drift toward plausible-looking names instead of fabricating arbitrarily. 
Moreover, we observe an cross-ecosystem effect, where some hallucinated names resemble packages from other language ecosystems.

\label{sec:RQ3}
\section{RQ3: Mitigation of Crate Hallucinations}

% \sout{Based on the findings of RQ1–RQ2, we design} 
In RQ3, we evaluate
two lightweight, language-sensitive interventions to mitigate hallucinations.
We keep the datasets and the detection pipeline unchanged and assess effectiveness solely by the change in CHR under controlled ablations, ensuring any improvement is attributable to the intervention rather than evaluation drift.
\subsection{Mitigation Strategies}

Our strategies focus on \textit{pre-processing} rather than post-processing, which is fragile in adversarial settings.
Attackers can obtain the name of a hallucinated package in advance, inject malicious code into the package, and publish it to code repositories, which makes post-processing methods ineffective~\cite{spracklen2025package}.
RQ2 also showed that hallucinations cluster around (i) standard-library modules miscast as external crates and (ii) near-miss names, suggesting that constraining dependency choices during generation is more promising.

Pre-processing strategies can generally be divided into two broad categories: Prompt Engineering and Model Development~\cite{tonmoy2024comprehensive}.
In this work, we concentrate on the former because it is deployment-friendly (no weight updates, no extra training cost).
Specifically, we employ two important methods of prompt engineering: \textit{Self-Refinement} and \textit{Retrieval-Augmented Generation (RAG).}

We chose RAG and self-refinement for the following reasons.
First, in a relatively less popular ecosystem like Rust, models may not reliably remember real crates and their intended usage.
RAG addresses this by incorporating an external source of crate names and descriptions and retrieving relevant evidence before generation, providing the model with grounded hints about existing crates rather than relying solely on its memory.
Second, code outputs in practice often go through review and iterative fixes.
Self-refinement automates this process by asking the model to audit its own output after generation and to revise the solution when issues are found.
Third, both approaches are training-free and require only standard inference access, without modifying model weights or accessing internal parameters.
%\\hc{What does it mean by black box}
This makes them portable across different models and convenient for controlled comparisons under the same evaluation metric.

\par\textbf{Self-Refinement.}
Self-Refinement is a prompt-level, inference-time procedure in which an LLM first produces an initial plan/answer, then re-evaluates it by running a second ``review'' prompt that asks the model to check for issues.
The model then revises the output based on this self-review, and this process can be repeated if needed~\cite{madaan2023self}.
We implement a verification-first, prompt-level self-refinement routine.
At generation time, the prompt instructs the model to detect standard-library functionality and explicitly qualify it via \texttt{std::}/\texttt{core::} modules.
When the model introduces an external crate, it is asked to perform a self-check on the exact crate name.
%\\hc{Why do we need to label these steps?}
If the model determines that the crate is a hallucination, it will regenerate the response with the instruction not to generate the crate again, repeating this process up to 3 times.

\par\textbf{Retrieval-Augmented Generation (RAG).}
RAG is a paradigm that augments a parametric generator with a non-parametric retriever to fetch task-relevant documents at inference time and generate conditions on the retrieved evidence, thereby improving factuality and mitigating hallucinations~\cite{lewis2020retrieval}.
Considering the balance between effectiveness and token cost,
we construct a dataset from \emph{crates.io} consisting of the top 10,000 crates ranked by recent downloads.
For each crate, we retain its canonical name and the description text provided on its \emph{crates.io} page.
When generating code, the database returns the top-five most relevant crates for the given task.
In parallel, we inject a fixed “STD list” that grounds common capabilities in the Rust standard library.
The list is seeded by the top-ten misused module names identified in RQ2,
and it is used to discourage unqualified references in generation.
When generating code, this list is appended to the prompt alongside the retrieved crate names, with an explicit policy that whenever the target capability is covered by the list, the model must use \texttt{std} and refrain from adding an external crate.

We implemented these mitigation strategies using the open-source models mentioned in Section \ref{sec:model_selection}, since they performed better than the commercial models during the experiments in RQ1.
Evaluation for commercial models is an important direction for future work, but proprietary APIs may introduce additional cost, access, and reproducibility constraints.
\subsection{Results}

Overall, both strategies reduce hallucinations for most evaluated models, but their effectiveness differs substantially.
As shown in Table \ref{tab:mitigation_performance},
self-refinement is consistently more effective than RAG, reducing crate hallucinations across all evaluated models.
On average, it lowers crate hallucinations by 2–3 percentage points, representing a 10–15\% relative reduction compared to each model’s original baseline.
In contrast, RAG provides only limited and less consistent improvements across the evaluated models.
This indicates that, under our current retrieval setup, RAG does not reliably suppress crate hallucinations and may be sensitive to retrieval quality and how well the retrieved evidence aligns with the model’s generation.
The results of the two lightweight strategies suggest that, while both approaches can help reduce hallucination rates for most models to some extent, crate hallucinations remain difficult to mitigate with simple, surface-level prompt interventions.
This observation is consistent with a prior study, which also reports that straightforward prompting or lightweight pre-processing alone is often insufficient to substantially eliminate package hallucinations~\cite{spracklen2025package}.
\begin{table}[t]
	\centering
	\caption{Performance of the Mitigation Strategies}
	\label{tab:mitigation_performance}
    \small
	\begin{tabular*}{\linewidth}{@{\extracolsep{\fill}}lrrr@{}}
		\toprule
		\textbf{Mitigation} & \textbf{DeepSeek} & \textbf{OpenCoder} & \textbf{Qwen2.5} \\
		\midrule
		No Mitigation                & 21.92\%           & 19.28\%            & 20.07\%          \\
        \midrule
		RAG                          & 20.0\% (-1.92\%)  & 18.32\% (-0.96\%)  & 20.41\% (+0.34\%)\\
		Self-Refinement              & 18.66\% (-3.26\%) & 17.06\%  (-2.22\%) & 17.76\% (-2.31\%)\\
		\bottomrule
	\end{tabular*}

\end{table}

\begin{rqblock}
	Both self-refinement and RAG reduce hallucination rates, but improvements are modest. Self-refinement is more effective, while RAG's gains are limited. 
\end{rqblock}

\textbf{Answer to RQ3.}
Both mitigation strategies reduce crate hallucinations, but self-refinement is more consistently effective than RAG under our setup.
Overall, lightweight prompting helps, but substantially mitigating Rust crate hallucinations likely requires stronger verification or tooling support beyond simple pre-processing.
% \jmrev{This motivates Rust-aware mitigation strategies that explicitly validate crate names, distinguish crates from modules, and align source-level references with Cargo-based dependency declarations.}

\section{Discussion}
% In this section, we discuss the implications of our empirical findings.
% We provide future-work suggestions on potential mitigations for Rust crate hallucinations, and we also discuss the limitations of our study and the factors that may pose threats to validity.

\subsection{Implications}

\par\textbf{Training Data Richness Drives Model Performance.}
Cassano et al.~\cite{cassano2024knowledge} categorize programming languages in LLM-based code generation into \textit{high-resource} and \textit{low-resource}, based on how much training data is available for each language.
Public statistics support this distinction: Rust is typically much smaller in scale than mainstream languages such as Python and JavaScript.
For example, the per-language statistics reported for The Stack dataset~\cite{kocetkov2022stack} show that Rust accounts for a substantially smaller amount of data (on the order of 40 GB) than Python (around 190 GB) and JavaScript (around 480 GB).
Under this categorization, Rust is therefore a typical low-resource language.
Prior work~\cite{orlanski2023measuring} also discussed how the distribution of programming languages in training data can affect model capability across languages: when training follows a natural distribution, LLMs perform worse on low-resource languages.
Consistent with this, our RQ1 results suggest that decoding-time parameter adjustments have limited impact on crate hallucination rates.
We further conjecture that model performance on Rust is driven more by training-data coverage than by model architecture or parameter settings, which helps explain why Rust can differ from high-resource languages such as Python and JavaScript in code generation behavior.

\par\textbf{Structure of Rust Crate Hallucinations.}
Our empirical results give the following key implications for understanding and mitigating crate hallucinations in Rust.
Crate hallucinations appear highly structured rather than purely arbitrary: many errors are tied to Rust-specific ecosystem and namespace conventions, such as mistaking standard-library or module paths as external crates, or producing near-miss crate names that resemble existing ones.

\par\textbf{Limited Impact of Decoding and Prompting.}
As mentioned above, changing decoding settings during generation is unlikely to substantially reduce hallucinations by itself, which motivates mitigation strategies that incorporate stronger training-time or external grounding mechanisms.
While lightweight prompt-based strategies can reduce hallucination rates for most models, the improvements are modest and not always consistent, indicating that package hallucination may not be substantially alleviated by simple prompting or pre-processing alone. More reliable mitigation likely requires tighter integration with toolchain-level verification.

%\yepang{We mentioned three points above. Can we organize them using three paragraphs or three itemized blocks?}
%\jm{revised. Merged point 1 into one paragraph.}

% \sout{\par\textbf{Directions for Mitigation.}
% For future work, we suggest exploring mitigation directions that integrate lightweight verification into Rust dependency recommendations. A practical direction is to equip coding assistants with registry-aware guardrails, so that any proposed crate is checked for existence.
% Given that many errors resemble plausible near-miss names, another promising line is confusion-aware recommendation, where tools flag suspiciously similar crate names and encourage users to confirm intent.
% Meanwhile, prior study in other language ecosystems suggests that fine-tuning can substantially reduce package hallucinations~\cite{spracklen2025package}.
% Therefore, future work may explore Rust-targeted fine-tuning as a promising direction to further mitigate this issue.}

\textbf{Directions for Rust-specific Mitigation.}
The modest improvements in RQ3 suggest that generic prompt-level methods are insufficient for fully mitigating crate hallucinations. 
We therefore view RAG and self-refinement as lightweight baselines rather than complete Rust-specific defenses. 
A more effective mitigation strategy should explicitly incorporate Rust’s crate/module namespace structure and Cargo-based dependency model.

Based on our findings in RQ2, a Rust-aware guardrail can be designed around three components. First, it should perform namespace-aware validation by distinguishing external crates from standard-library modules, in-crate modules, and rustc crates. 
This directly targets module-crate confusion, where valid Rust modules are incorrectly treated as external dependencies. 
Second, it should perform registry-aware and near-miss checking against \emph{crates.io} and known aliases. 
Since many hallucinated names are close variants of real crates, suspicious names can be flagged or corrected using edit-distance and naming-convention checks. Third, it should be Cargo-aware: source-level references should be cross-checked with \texttt{Cargo.toml}, because Rust dependencies are resolved through the manifest rather than by source imports alone.
Overall, effective crate-hallucination mitigation should combine LLM generation with Rust-aware verification. 
Instead of relying only on retrieved context, future coding assistants should validate whether a generated identifier is a crate, module, alias, or invalid dependency, and then provide feedback for correction or regeneration.

\subsection{Threats to Validity}
The LLM  evolves rapidly, so hallucination rates and model differences observed within our evaluation window may not directly generalize to newer models or more tool-integrated assistants.
In addition, our results are based on a fixed set of models and a Rust-specific prompt dataset constructed from selected sources.
While these sources provide complementary coverage, they may not fully represent all Rust development contexts.
To mitigate this threat, we evaluate multiple model families with different sizes, which adds confidence to our generalizability.
For the synthetic task subset, it is built by Qwen2.5-32B. This may introduce distribution bias for model families related to the generator. However, in our aggregate results, we did not observe an obvious favorable outlier pattern for the Qwen family.
Our metric labels an extracted crate name as hallucinated if it is absent from the reference list.
This may introduce false positives when a model mentions an alias, an outdated name, or a non-crate identifier that is incorrectly parsed as a crate.
False positives may also occur when a model refers to private crates that are not indexed by public registries.
To reduce these risks, we use a consistent extraction pipeline and apply the same detection rule across all settings.
Although this does not eliminate mislabeling, it ensures that comparisons across models and mitigation settings are made under the same operational definition and data-processing procedure.
Our study does not directly measure downstream consequences such as build failure rates, developer repair effort, or the exploitability of hallucinated crate names. Therefore, CHR should be interpreted as a source-level indicator of dependency confusion rather than direct evidence that a hallucinated reference will necessarily lead to package installation or supply-chain compromise.

\label{sec:discussion}

\label{sec:related}
\section{Related Work}
Our work draws inspiration from two lines of related work.
We discuss them in this section.

\subsection{Hallucination in Code Generation}
Recent studies have begun to systematically characterize hallucinations in LLM-based code generation. 
Liu et al.~\cite{liu2024exploring} provide an early taxonomy by sampling tasks from HumanEval~\cite{chen2021evaluating} and DS-1000~\cite{lai2023ds}.
They categorize code hallucinations into three major types: intent-conflicting, context deviation, and knowledge-conflicting.
Extending beyond function-level tasks, Zhang et al.~\cite{zhang2025llm} argue that practical development settings introduce richer failure modes and propose a taxonomy for repository-level generation that emphasizes requirement, factual, and project-context conflicts.
Complementary perspectives define hallucinations via execution outcomes: Tian et al.~\cite{tian2025codehalu} summarize runtime-oriented error categories (e.g., mapping, naming, resource, and logic failures) and introduce \textsc{CodeHaluEval}.
Agarwal et al.~\cite{agarwal2025codemirage} organize hallucinations by defect types and construct \textsc{CodeMirage}.
Finally, a systematic review by Gao et al.~\cite{gao2025systematic} synthesizes definitions and proposes a unified framework that highlights knowledge, functional, and environment/dependency-related hallucinations, offering a consolidated view of the field.

Overall, taxonomy-driven studies provide a shared vocabulary and evaluation basis for systematically characterizing code hallucinations across tasks and models. 
% They also make hallucination patterns more actionable by linking recurrent failure types to targeted mitigation and benchmarking.

\subsection{Package Hallucination}
Recent work has formalized package hallucination as a supply-chain–relevant phenomenon in LLM-assisted coding.
Spracklen et al.~\cite{spracklen2025package} and Krishna et al.~\cite{krishna2025importing} provided foundational empirical studies, focusing on mainstream languages such as Python and JavaScript. 
They identified package hallucination as a prevalent phenomenon, analyzed behavioral traits, and evaluated mitigation strategies. Both studies found that hallucination rates and patterns depend on the model, its size, and language, but left open questions regarding less-studied ecosystems.
Beyond empirical studies, Zhao et al. designed \textsc{HFuzzer}~\cite{zhao2025hfuzzer}, a fuzzing technique for systematically testing LLMs for Python package hallucinations. This approach triggered more unique hallucinations and generated more diverse coding tasks than previous methods such as \textsc{GPTFuzzer-A}~\cite{yu2309gptfuzzer}.

Together, these studies established a clear research trajectory, from characterizing package hallucinations to advancing testing techniques. 
Extending this line of work, we investigate package hallucination in Rust, conduct a comprehensive evaluation, characterize its Rust-specific properties, and compare our findings with those from more widely studied mainstream languages.

% \yepang{Related work section is rather short. It seems that we only discussed four papers. Please include more. You may consider organizing the existing studies into several different topics and discuss each topic using a subsection.}

\label{sec:conclusion}
\section{Conclusion}
In this work, we study crate hallucinations in LLM-generated Rust code.
We constructed a multi-source coding task dataset combining data from \textsc{Stack Overflow}, \textsc{GitHub}, and LLM-generated tasks, which allowed us to cover a wide range of Rust coding tasks and dependency scenarios.
We then evaluated Rust crate hallucinations across six LLM models to assess the impact of model family, model size, and decoding parameters. 
We found that crate hallucinations remain consistent across those parameters mentioned above.
Our results show that Rust crate hallucinations are highly structured rather than random, with many errors arising from Rust-specific conventions.
%\yepang{You may also wish to highlight the effect of model family, size, and decoding parameters in conclusion.}
While lightweight mitigation strategies such as self-refinement and RAG were found to reduce hallucination rates, the improvements were modest and not always consistent.
We hope our findings can provide insights for future research on package hallucinations, enabling follow-up work to propose effective mitigation strategies specifically tailored to the Rust language.

\section{Data Availability}
\label{sec:data_acountbility}

% \sout{The datasets used in this study are available via an anonymous private link:
% \url{https://figshare.com/s/f19fe17afb291ad386fa}.
% We will release a public, archival version upon acceptance.}

The task datasets, prompt templates, crate extraction scripts, evaluation scripts, mitigation scripts, and result files in this study are available via this link:
\url{https://doi.org/10.6084/m9.figshare.32233767}.

\begin{acks}
This work is supported by
the National Natural Science Foundation of China (Grant No.\ 62372219) and the Fundamental Research Funds for the Central Universities (No.\ 63261089).
\end{acks}

% Bibliography - using ACM reference format
\bibliographystyle{ACM-Reference-Format}
\bibliography{reference}

\end{document}